\documentclass[aps,twocolumn,prd]{revtex4}

\def\be{\begin{equation}}
\def\ee{\end{equation}}
\def\bea{\begin{eqnarray}}
\def\eea{\end{eqnarray}}
\def\ba{\begin{array}} 
\def\ea{\end{array}}
\def\bc{\begin{center}}
\def\ec{\end{center}}

\def\ghost#1{}

\def\sl{\!\!\!/}

\def\simge{\mathrel{%
   \rlap{\raise 0.511ex \hbox{$>$}}{\lower 0.511ex \hbox{$\sim$}}}}
\def\simle{\mathrel{
   \rlap{\raise 0.511ex \hbox{$<$}}{\lower 0.511ex \hbox{$\sim$}}}}

\def\physl#1#2#3{Phys. Lett. B#1, #2 (#3)}
\def\physrev#1#2#3{Phys. Rev. D#1, #2 (#3)}
\def\prl#1#2#3{Phys. Rev. Lett. #1, #2 (#3)}
\def\nucl#1#2#3{Nucl. Phys. B#1, #2 (#3)}

\begin{document}

\title{Light \,spin-$\bf \frac{1}{2}$\, or \,spin-0\, Dark Matter particles
\medskip \\ }
\author{P. Fayet\,$^1$
\vspace{3mm}
\\
$^1$ Laboratoire de Physique Th\'eorique de l'ENS
{\small \ (UMR 8549 CNRS)} \\
 24 rue Lhomond, 75231 Paris Cedex 05, France}
\date{March 20, 2004}

\begin{abstract}
{
We recall and precise how light spin-0 particles could be acceptable
Dark Matter candidates, and extend this analysis to
spin-$1/2$ particles. We evaluate the (rather large) annihilation cross sections
required, and show how they may be induced by a new light neutral spin-1 boson $U$.
If this one is vectorially coupled to matter particles,
the \,(spin-$1/2$ \,or spin-0)\, Dark Matter annihilation cross section into $\,e^+e^-$
automatically includes a $v_{dm}^{\,2}$ suppression factor at threshold,
as desirable to avoid an excessive production
of $\,\gamma\,$ rays from residual Dark Matter annihilations.
We also relate Dark Matter annihilations
with production cross sections in $\,e^+e^-$ scatterings.
Annihilation cross sections of spin-$1/2$
and spin-0 Dark Matter particles are given
by exactly the same expressions.
Just as for spin-0, light spin-$1/2$ Dark Matter particles
annihilating into $\,e^+e^-\,$
could be responsible for the bright \,511 keV
$\,\gamma$ ray line observed by INTEGRAL from the galactic bulge.
}
\end{abstract}


\maketitle

\section{Introduction}

\label{sec:intro}

Weakly-interacting massive neutral particles,
taken as possible Dark Matter candidates, should not be too light,
otherwise they would not have been able to annihilate sufficiently.
Weakly-interacting heavy neutrinos
would have had to be  heavier than about 2 GeV, for example
(to get $\,\Omega_\nu \,h^2\simle 1$) \cite{leew}.

\vspace{2mm}

Supersymmetric extensions of the Standard Model
naturally provide such weakly-interacting neutral particles,
stable as a result of $R$-parity conservation
(with $R_p=(-1)^{2S}\,(-1)^{(3B+L)}\,$) \cite{ssm}.
Spin-$\frac{1}{2}$ photinos, or more generally neutralinos,
with cross sections roughly of weak-interaction order
when the exchanged squarks and sleptons are
$\sim m_W$  \cite{sigmaphot}, \,should be heavier
than a few GeV's (for light sfermion masses) at least
to annihilate sufficiently, this bound increasing
with the exchanged sfer\-mion masses $m_{\tilde q,\,\tilde l}$
\cite{goldberg,revuedm}.
Given the still unsuccessful hunt for superpartners, in particular at LEP,
the lightest neutralino (LSP) of Supersymmetric extensions of the Standard Model
is now generally believed to be heavier than about $\,\sim 30\,$ GeV.

\vspace{2mm}

Then, how could a light (annihilating) Dark Matter particle possibly exist?
At first it should have {\it no significant direct coupling to the $Z$ boson},
otherwise it would have been produced in $Z$ decays at LEP.
Despite that, it would have to {\it annihilate sufficiently\,}
-- and in fact, much more strongly than through ordinary weak interactions
-- otherwise its relic energy density would be too high\,!
Can this happen at all, and what could then be the new interactions
responsible for Light Dark Matter annihilations\,?

\vspace{2mm}
We have explored in \cite{bf} under which conditions
a light \hbox{spin-0} particle
could be a viable Dark Matter candidate.
Two different situations have been exhibited,
in which the new interactions responsible for the annihilations
are due to non-chiral couplings with exchanged heavy fermions such as mirror fermions
(case I), in (supersymmetric) theories somewhat reminiscent of $N=2$ extended supersymmetry
and/or higher-dimensional theories \cite{mirror}.
Or, such interactions may be mediated by a new neutral spin-1 gauge boson $U$ (case II),
similar to the one, light and very weakly coupled,
introduced long ago \cite{pfu}.
It is also desirable that the Dark Matter pair annihilation cross section
into $e^+e^-$
has a $v_{dm}^{\,2}$ suppression factor, so as to avoid an excessive production
of $\,\gamma\,$ rays originating from the residual annihilations of Dark Matter
particles (if lighter than $\sim$ 100 MeV) \cite{bes}.
This is naturally the case when these spin-0 particle annihilations result from the
virtual production of a new spin-1 $\,U$ boson.

\vspace{2mm}

We shall show here that, while the first situation (I)
is specific of spin-0 particles, the second (II) is not,
and could apply to spin-$\frac{1}{2}$ as well as to spin-0 particles.
One crucial feature is that
the new interactions mediated by the $U$ boson should actually be
``not-so-weak'' (at lower energies and relatively to weak interactions)
\,-- i.e. $<\!\sigma_{ann} \,v_{rel}/c\!> \ \,\approx $ \,a few (up to $\,\approx$ 10) picobarns --\,
so as to ensure for sufficient annihilations
of light Dark Matter particles, whatever their spin.
More precisely, the new $\,U$-mediated Dark-Matter/Matter interactions
will be {\it stronger than ordinary weak interactions at lower energies\,}
--\, but {\it weaker at higher energies}, \,at which they are damped by
$\,U$ propagator effects.
The smallness of the $U$ couplings to ordinary matter,
as compared to $\,e$, by several orders of magnitude,
\,then accounts for the fact that
these particles have not been observed yet.

\vspace{2mm}
A second essential feature is that the annihilation cross sections of
such spin-$\frac{1}{2}$
Dark Matter particles into fermion pairs $f\bar f$
through the exchanges of a new neutral spin-1 $U$ boson,
will, also in this case, have the desired  $v_{dm}^{\,2}$ suppression factor
at threshold, provided the $U$ boson is {\it \,vectorially\,} coupled to
{\it \,matter\,} fermions,
as is in any case necessary to avoid a problematic axionlike behavior
of its longitudinal polarization state \cite{pfu}.

\vspace{2mm}
Indeed, as we shall see, the annihilation, at threshold, of a
$\,C=+\,$ state (made of two Majorana particles $\,\chi$,
with $\,J=L=S=0$) \,into a $\,f\bar f\,$ final state
with $\,C'=(-)^{(L'+S')}= +\,$, \,through a $\,C$-violating interaction
(axial $\,\chi\,$ current {\footnotesize \,times\,} vector $f$ current),
is forbidden by charge conjugation.
This ensures that the annihilation cross section
$\,\sigma_{ann}\,v_{rel}\,(\chi\chi\to e^+e^-)\,$
has the appropriate $\,\propto\,v_{dm}^{\,2}\,$ behavior,
automatically suppressing (by a factor $\,\approx10^{-5}\,$)
the late annihilations of non-relativistic relic Dark Matter particles.

\vspace{2mm}
Furthermore, the annihilation cross sections
of \hbox{spin-$\frac{1}{2}$} and spin-0 Dark Matter particles will, in this case,
be given by {\it \,exactly the same expressions}.
\hbox{Spin-$\frac{1}{2}$} particles then turn out to be acceptable
Light Dark Matter (LDM) candidates, as well as spin-0 particles.
In particular, their annihilations into $\,e^+e^-$ pairs
could lead to a $\,\gamma\,$ ray signature from the galactic center at low energy,
as indicated for spin-0 particles in \cite{bf} (before the observations of \cite{integral}).
Just as the latter, they could be responsible for the bright \,511\, keV
\,$\gamma\,$ ray line recently observed by the INTEGRAL satellite from the galactic bulge
\cite{integral,betal,bfsi}. \,Other effects of such light Dark Matter particles,
on nucleosynthesis and energy transfer in stars, were discussed very recently in
\cite{sr}.

\section {Dark Matter decoupling \break and relic density}
\label{sec:relic}

The interactions responsible for the pair annihilations
of spin-$\frac{1}{2}$ Dark Matter particles $\chi$
(such as heavy neutrinos or neutralinos, ...)
may be written, in the local limit approximation,
as effective four-fermion interactions
$\,{\cal L}\,\approx \,G\ \,\bar \chi ... \chi\ \ \bar f ... f\,$.
The corresponding annihilation cross sections,
proportional to $\,G^2$, \,scale essentially like $\,m_{dm}^{\,2},$
$\,m_{dm}$ being the Dark Matter particle mass.
Such {\it \,fermionic\,} particles annihilating through exchanges of heavy
{\it \,bosons\,} of masses $\,\simge m_W$ cannot be light
(in a perturbative theory), since their annihilation cross sections
would be too small.

\vspace{1mm}
To estimate what annihilation cross sections are actually needed
for a correct relic abundance of light Dark Matter particles
(corresponding to $\,\Omega_{dm}\,h^2\simeq .1$), \,we
express that the annihilation rate
$\,\Gamma =n_{dm}\!<\!\sigma_{ann}\,v_{rel}\!>\,$
and expansion rate $\,H\,$
are approximately equal when the Dark Matter annihilation reactions freeze out.
This occurs at a temperature $\,T_F=m_{dm}/x_F$,
\,with $\,x_F$ roughly between $\,\simeq \,16\,$ to $\,\simeq \,23 \,$ for a 1 MeV to 1 GeV particle
(cf. Appendix \ref{app:xf}).

\vspace{3mm}
{\Large $\bullet$}\ \ For {\bf \,10 MeV  $\boldmath \simle m_{dm}\simle 1$ GeV}, \,the freeze-out occurs at $T_F$
(with, roughly, \,.6 MeV $\,\simle \,T_F \simle \,50\,$ MeV) after muons have annihilated
(most of them at least), but not electrons yet. The effective number of degrees of freedom is then $\,g_*\simeq 43/4$.
The surviving particles get diluted by
the expansion of the Universe, proportionally to $\,T^3$,
\,with an extra factor $\,4/11\,$ corresponding to the subsequent annihilation
of $\,e^+e^-$ pairs into photons, so that the relic density of Dark Matter particles
may now be expressed as
\,\footnote{For heavier particles decoupling before muon annihilations, there is an extra dilution factor which may be written as $\,\frac{43}{4}/g'_*\,$, \,while for lighter particles decoupling after $\,e^+e^-$ annihilations the factor \,4/11\, is absent.}
\be
\label{densite0}
n_{\circ \,dm}\ \,=\ \frac{4}{11}\ \,\frac{T_{\circ\gamma}^{\ 3}}{T_F^{\,3}}\ \ n_{dm}\ \ ,
\ee
$T_{\circ\gamma}\simeq \,2.725 \ \hbox{K}\, \simeq \,2.35\ 10^{-13}\ \hbox{GeV}\
\,(\simeq\,11.9\ \hbox{cm}^{-1})\,$ being the present photon temperature.
We shall denote by $\,N_{\circ\,dm} = (2) \ n_{\circ\,dm}\,$ the total density of
Dark Matter (particles~+~antiparticles), with the factor 2 present only in the case of
non-self-conjugate Dark Matter particles.

\vspace{2mm}

The resulting freeze-out equation $\ \Gamma\,\simeq\,H\,$, \,i.e.
\be
\label{gamma=h0}
n_{\circ\,{dm}}\ \frac{11}{4}\ \frac{T_F^{\,3}}{T_{\circ\gamma}^{\,3}}\
<\sigma_{ann}\,v_{rel}>\ \,\simeq\
1.66\ \sqrt{\hbox{\small $g_*\!=\frac{43}{4}$}}\ \ \frac{T_F^{\,2}}{m_{Pl}}\ ,
\ee
sufficient as a first approximation,
fixes the relic energy density
\be
\label{rhodm0}
\ba{ccl}
\rho_{dm}  &\simeq&
\ (2)\ \,n_{\circ\,{dm}}\ m_{dm}\,\simeq \ (2) \ x_F\
n_{\circ\,{dm}}\,T_F
\vspace{2mm}\\
&\simeq& \displaystyle  (2)\ \,\frac{4}{11}\ \,1.66\ \sqrt{\hbox{\small $g_*\!=\frac{43}{4}$}}
\ \, x_F\ \frac{T_{\circ\gamma}^{\,3}}{m_{Pl}}\ \
\frac{1}{<\sigma_{ann}\,v_{rel}>}
\vspace{2mm}\\
&\simeq& \displaystyle  (2)\ \,\frac{x_F}{20}\ \ \
\frac{4.2\ 10^{-56}\ {\rm GeV}^2}{<\sigma_{ann}\,v_{rel}>}\ \ .
\ea
\ee
Dividing by the critical density $\,\rho_c/h_\circ^{\,2}\simeq 1.054\ 10^{-5}$\ GeV/cm$^3$
\,(times $\,h\!\!\!/ \,\simeq \,6.58 \ 10^{-25}$ GeV.\,s) \,to get the density ratio $\,\Omega_{dm}\,h^2$, \,we find
\be
\label{omegah20}
\ba{ccl}
\displaystyle
\frac{\Omega_{dm}\ h^2}{.1} & \simeq & \displaystyle \frac{x_F}{20}\ \
(2)\ \frac{6\ \,10^{-26}\ {\rm cm}^3/{\rm s}}{<\sigma_{ann} \,v_{rel}>}\
\vspace{2mm} \\
&\simeq & \displaystyle
(2)\ \ \frac{x_F}{20}\ \
\frac{2\ \,10^{-36}\ {\rm cm}^2}{<\sigma_{ann} \,v_{rel}/c>}\ \ ,
\ea
\ee

\noindent
with the extra factor 2 present in the case of non self-conjugate particles.

\vspace{4mm}

More precisely, there is also,
from an approximate solution of the Boltzmann equation,
an expected increase of the required cross section by a factor $\,\approx 2$,
\,for \linebreak
$\,<\!\sigma_{ann}v_{rel}\!>\,$ \,behaving at threshold like
$\,v_{dm}^{\,2}\,$, \,as compared to a constant \cite{kt}
\footnote{For comparison with other results, our equation (\ref{rhodm0})
fixing $\, \rho_{dm},\,$
with the extra dilution factor $\frac{43}{4}/g'_*\,$ relevant for heavier Dark Matter particles included, may be rewritten, using $\,T_{\circ\,\gamma}^{\,3}\simeq \,1685\,$ cm$^{-3}$, \,as
\vspace{-1mm}
$$
\ \ \ \ \ \ \rho_{dm}\ \approx \ (2)\ \,
1.09\ 10^4 \ \ \frac{x_F}{g'_*/\sqrt g_*}\
\frac{1}{m_{Pl}\,<\sigma v>}\ \ \hbox{cm}^{-3}\ \ ,
$$
\vspace{-1mm}
and, dividing by $\,\rho_c/h^2\,$,
$$
\ \ \ \ \ \ \ \Omega_{dm}\ h^2  \,\simeq\  (2)\ \
\frac{x_F}{g'_*/\sqrt{\,g_*}}\ \
\frac{1.04\ \,10^9\ }{m_{Pl}\,\times\, (1\ \hbox{GeV})\ \,<\sigma v>}\ ,
$$
as in \cite{kt}, for a velocity-independent $\,<\sigma_{ann}\,v_{rel}>\,$.
}.

\vspace{2mm}
Indeed the
later annihilations that would still occur below the temperature $\,T_F$ given by eq.\,(\ref{gamma=h0})
are further inhibited by this $\ v_{dm}^{\,2}\,$ factor, preventing the Dark Matter density from reaching the equilibrium value corresponding to this $\,T_F$, as
it would be given by eqs.\,(\ref{gamma=h0}) to (\ref{omegah20}).
Altogether, obtaining the right amount of Dark Matter ($\,\Omega_{dm}\,h^2\simeq \,0.1\,$)
\,requires typically
\be
\label{sigv}
<\sigma_{ann}v_{rel}/c >
\ \simeq\ (2)\ \,(\,4\,\ \hbox{or} \,\ 2\,)\ \hbox{pb}\ \ ,
\ee

\vspace{1mm}
\noindent
depending whether $\,<\sigma_{ann}\,v_{rel}/c>\,$  behaves like
$\,v_{dm}^{\,2}$ ($\,\simeq$ \,(2)\ 4  pb, the most interesting case for us here),
\,or as a constant ($\,\simeq $\, (2)\ 2 pb)\,\footnote{These formulas,
which apply directly for $\,m_{dm}\,$ in the \hbox{10 MeV} -- 1 GeV range,
can be immediately extended to heavier particles by including the extra factor
$\ \frac{\sqrt{g_*\,(43/4)}}{g'_*}$ \,(usually the same as
$\,\sqrt{(43/4)/g_*}$\ ), \,and using the appropriate (slightly larger) value of $\,x_F$.
\,For comparison, in the familiar case of {\it \,heavy\,} Dark Matter particles,
the significantly larger value of $g_*$
brings the required cross sections
$\,<\sigma_{ann}\,v_{rel}>\,$ close to about $\,(2)\ \,3\ 10^{-26}\ \hbox{cm}^3/$s \,(as compared to (\ref{omegah20})), \,in the case of $\,v_{dm}\,$-independent cross sections \,(resp. $\,(2)\ 6\ 10^{-26}\ \hbox{cm}^3/$s \,when they are $\,\propto\, v_{dm}^{\,2}$).
},
the factor \,2\, being associated with non self-conjugate Dark Matter particles.

\vspace{3mm}
{\Large $\bullet$}\ \,Let us now consider {\it lighter Dark matter particles}
(say {\bf $\boldmath m_{dm}\simle 10$ MeV}),
actually the most interesting situation. At first,
for particles lighter than about 2 to 3 MeV,
that would decouple (at $\,T_F=m_{dm}/x_F \simle .15$ MeV)
\,after most electrons have annihilated,
the dilution factor of $\,4/11\,$ is no longer present in eqs.\,(\ref{densite0}-\ref{rhodm0}).
In addition, the $\,g_*\,$ at Dark Matter freeze-out is no longer 43/4\, when electrons have disappeared. It may in fact be expressed in terms of the neutrino temperature as
\be
g_*\ \simeq\ 2\ +\ \frac{7}{8}\ \,(2\times 3)\ \ \left.\frac{T_\nu^{\ 4}}{T^4}\right. _F\ \ ,
\ee
which would be $\,\simeq 3.36\,$ according to the standard model, where
$\,T_\nu/T \simeq (4/11)^{1/3}$ as an effect of electron annihilations.

\vspace{2mm}
The neutrino contribution to $\,g_*\,$, \,however, is then no longer the same as in the standard model.
Indeed Dark Matter particles annihilating after neutrino decouple, at $\,T\approx 3.5\,$ MeV for $\,\nu_\mu,\,\nu_\tau $ or $\,\approx 2\,$ MeV for $\,\nu_e$ (assuming neutrino interactions with Dark Matter do not keep them longer in thermal equilibrium with photons, as discussed in \cite{sr}) would also heat up the photon gas
as compared to neutrinos, so that the resulting neutrino temperature would be less
than the usual $\,(4/11)^{1/3}$, \,resulting in {\it a lower contribution of neutrinos\,} to the
$\,g_*\,$ at $T_F$\
than in the standard model (down to \,.71\, to \,.94 instead of \,1.36), due to Dark Matter annihilations themselves\,\footnote{
Once Dark Matter particles and electrons have annihilated,
the neutrino temperature gets related to the photon temperature by
\vspace{-1mm} \\
$$
\frac{T_\nu^{\,3}}{T^{\,3}}\ \simeq\
\frac{g_{*\gamma}=2}
{(g_{*\,e+\gamma}=\frac{11}{2})\,+\, g_{*\,dm}}\ \simeq\ \frac{8}{29,\ 36 \ \hbox{\,or}\ 30}\ \ ,
$$
with
$\ g_{*\,dm} \,=\, \frac{7}{4},\ \frac{7}{2}\ \,\hbox{or}\ 2\,$
for a Majorana, Dirac or complex spin-0 particle, respectively.
This leads to a total $\,g_{*\,F}\,\simeq\, 2.94,\ 2.71\ $ or \,2.90, respectively, instead of
3.36 in the standard model.
}\,!

\vspace{2mm}

Such a phenomenon, if a significant fraction of Dark Matter annihilations were to occur
after neutrino decoupling
but before the neutron/proton ratio freezes out,
could have important implications,
potentially allowing for less primordial helium than in the standard model
(or conversely allowing for new species, e.g. additional light inos, etc.,
to contribute to the expansion rate in a way which would otherwise have been forbidden).
This effect, that we found qualitatively, is discussed in detail in \cite{sr},
as well as, more generally, the effects of light Dark Matter particles on
the big bang nucleosynthesis, with the conclusion that light masses
$\,m_{dm}\simle\,2\,$ MeV are disfavored as they would severely disturb the BBN concordance.

\vspace{2mm}
Given INTEGRAL results, we tend to favor light Dark Matter masses just above this value,
so as to maximize the number of $\,e^+$ produced for a given Dark Matter energy density;
and, also, to avoid these $\,e^+$ from Dark Matter annihilations (produced with an energy close to $\,m_{dm}$) \,having too much energy dissipated in $\gamma$ rays
(as it would happen for a not-so-small $\,m_{dm}$) \,before they can form positronium and annihilate,
leading to the bright 511 keV $\,\gamma$ ray line.

\vspace{3mm}
Ignoring for a moment this mass restriction as we discuss relic abundances,
for a particle in the $\,\simeq \,\frac{1}{2}\,-\, 2\,$ MeV mass range,
the required cross section gets increased
(from the absence of the 4/11 dilution factor and the lower value of
$\,g_*\,,$ \,and $\,x_f$) \,by a factor
$\,\frac{11}{4}\ \sqrt{g_*/(43/4)}\ \,(\frac{x_F}{20}\simeq .85)\,\simeq \,1.2\,$,
\,as compared to a $\,\approx$ $\,100$ MeV particle, a rather moderate increase of about 20\,\%.
No spectacular difference is then expected  when the mass grows from 2\, to 10 MeV,
the effects of the 4/11 dilution factor and of the larger $g_*$ approaching \,43/4\, getting progressively reestablished.

\vspace{6mm}
{\Large $\bullet$}\ \,Altogether {\it \,for cross sections behaving like} $v_{dm}^{\,2}\,$,
the required annihilation cross sections at freeze-out
$\,\hbox{$<\sigma_{ann}\,v_{rel}/c>$}\ $
are of the order of \,4 to 5\,  picobarns for a self-conjugate (Majorana) Dark Matter particle,
or \,8 to 10\, picobarns  for a non self-conjugate one, e.g. a complex scalar, as
summarized above in Table \ref{sigma}. \,(We do not consider real self-conjugate spin-0 particles, as Bose statistics does not allow for
the desired $P$-wave annihilation.)

\begin{table}
\caption{\label{sigma}
Estimates of the annihilation cross sections \hbox{$\,<\sigma_{ann}\,v_{rel}/c>\,$} at freeze out
\,required for a correct relic abundance \,(\,$\Omega_{dm}\,h^2\simeq\,.1\,$).}

\vspace{3mm}

\begin{tabular}{|c|c|c|}

\hline && \vspace{-2mm} \\
\ \ Spin-$\frac{1}{2}$ Majorana \ \   &\ \  Spin-$\frac{1}{2}$ Dirac\ \  & \ \ Spin-0 \ \
 \\
{\small ($\chi$)} & {\small ($\psi$)} &
{\small \ \ \, ($\varphi$ complex)} \ \ \, \\
\vspace{-3mm} && \\ \hline
\vspace{-2mm}&& \\
4 -- 5\ pb & 8 -- 10\ pb & \ 8 -- 10\ pb
\vspace{-1mm}\\
&& \\ \hline
\end{tabular}
\end{table}

\vspace{2mm}

This corresponds roughly, for present annihilations of residual
Dark Matter particles
having a velocity $\,v_{dm}\approx 3\ 10^{-3}$ the velocity
at freeze-out in the primordial Universe ($\,\simeq \,.4\ c\,$),
to annihilation cross sections
\be
<\sigma_{ann}\,v_{rel}/c>_\circ\ \  \simeq \ \,(\,4\,\ \hbox{to}\ 10\,)\ 10^{-5}\
\hbox{pb}\ \ ,
\ee
respectively. Although large, this is indeed the right order of magnitude for
light Dark Matter particle (in the $\,\simeq\,$ MeV range)
annihilations to be at the origin of the \,511 keV $\gamma$
ray signal observed by INTEGRAL from the galactic bulge\,\cite{integral, betal}
\footnote{With a measured flux of about $\,10^{-3}$ photon cm$^{-2}$ s$^{-1}$,
the required cross section may be estimated $\propto 10^{-4}$ pb,
\,times $\,(m_{dm}/\hbox{1\ MeV})^2$, \,depending on the density distribution
within the galactic bulge. Note also that the rate of annihilation events
is proportional to
$\,\frac{1}{2}\ N_{dm}^{\,2} $ \hbox{$<\!\sigma_{ann} v_{rel}\!>\,$} for
self-conjugate Dark matter particles, and to
$\,(\frac{N_{dm}}{2})^2 <\!\sigma_{ann}\,v_{rel}\!>\,$,
\,for non self-conjugate ones (the number densities of
particles and antiparticles being both $\frac{N_{dm}}{2}$).
The required cross sections are thus, also in this case, twice as large
for non self-conjugate particles as for self-conjugate ones.}.

\vspace{2.5mm}
In any case, since
$\,G_{\!F}^{\,2}\,(1\,\hbox{GeV})^2\,/2\,\pi
\simeq .8\ 10^{-38}$ cm$^2$,
cross sections (at freeze-out) of weak interaction order are,
for light masses $\,m_{dm} \ll$ GeV,
\,by far too small for a correct relic abundance.
Significantly larger annihilation cross sections are needed,
requiring new types of interactions,
as discussed in \cite{bf}.

\section{Annihilations through heavy fermion ex\-chan\-ges}

\vspace{-2mm}

In case (I) (cf.~Introduction), one arranges,
for spin-0 Dark Matter particles,
to have  annihilation cross sections
behaving as the inverse of the (large) {\it \,squared masses\,}
of exchanged {\it \,fermions}
(rather than the 4$^{\rm th}\,$ power of exchanged boson masses).
Spin-0 Dark Matter particles \,($\varphi$)\,
are taken to have Yukawa interactions
coupling ordinary quarks and leptons $f$ to heavy fermions $F$
such as {\it \,mirror fermions\,}\,\footnote{Furthermore in theories inspired
from $N=2$ extended supersymmetry, or higher-dimensional
theories \cite{mirror}, the \hbox{spin-0\,} field $\,\varphi\,$ may be a spin-0 photon field,
or companion of the weak hypercharge gauge field $\,B^\mu$,
possibly described by the extra 5$^{\rm th}$ and 6$^{\rm th}$
components of the corresponding 6d-gauge fields, $\,A^{\hat\mu}$ or $B^{\hat\mu}$...
}.
The low-energy effective Lagrangian density responsible for their
pair-annihilation into $\,f\bar f\,$
may be written as
\be
\label{leff2}
{\cal L}\ \ \approx \ \
\frac{C_l\,C_r}{m_F}\ \ \varphi^*\varphi\ \ \overline{f_R}\,f_L\,
+\ \hbox{h.c.}\ \ ,
\ee
where $C_l$ and $C_r$ denote the Yukawa couplings of the \hbox{spin-0}
Dark Matter particles to the left-handed and right-handed fermion fields,
respectively. The resulting annihilation cross section at threshold
is of the type
\be
\label{sigmamf}
\sigma_{ann}\,v_{rel}\ \ \approx\ \ \frac{C_l^2\,C_r^2}{\pi\ m_F^2}\ \ ,
\ee

\vspace{-1.5mm}
\noindent
in the case of {\it \,non-chiral\,} couplings (i.e. for $\,C_l\,C_r\,\neq0$).
\linebreak
This cross section, largely independent of the Dark Matter mass,
can be quite significant
{\it \,even for light \hbox{spin-0\ } Dark Matter particles} \footnote{For example,
$\,\alpha^2/\,(100\ \hbox{GeV})^2 \simeq 2\ 10^{-36}\ \hbox{cm}^2,$
which shows that appropriate values of
$\,<\!\sigma_{ann}\,v_{rel}/c\!>\ \approx$ \hbox{(2 to 5)} $10^{-36}$ cm$^2$\,
may easily be obtained, from the exchanges of heavy
fermions $F\,$ in the \,$\sim$ 100 -- 1000 GeV mass range,
with Yukawa couplings $\,C_l\,$ and $\,C_r$ remaining
in the perturbative regime.}.

\vspace{1mm}

However, in the absence of a $P$-wave suppression factor
proportional to $\,v_{dm}^2$
(since the couplings (\ref{leff2})
involve fermion fields of {\it \,both\,} chiralities,
allowing for non-vanishing $S$-wave annihilations into $f\bar f$)\,\footnote{The effective
Lagrangian density
(\ref{leff2}) is invariant under Charge Conjugation,
allowing for a non-vanishing annihilation cross section
at threshold of a $\,\varphi\bar \varphi\,$ initial state
with $C=(-)^L\!=+$ \,into a $f\bar f$ state
with $\,C'=(-)^{(L'+S')}\!=+$.},
one runs the risk, at least for lighter Dark Matter particles
($\,\simle 100$ MeV$/c^2$) of too much $\gamma$ ray production due to
residual annihilations of Dark Matter particles \cite{bes}
(unless there is an asymmetry between Dark Matter particles and antiparticles).

\vspace{1mm}

It is thus preferable to consider annihilations induced through the virtual production
of a new light
neutral spin-1 gauge boson $\,U$ (case II). We can then both
get an appropriate relic abundance,
together with the desired  $\,v_{dm}^{\,2}\,$ suppression factor
in the annihilation cross sections.
We shall point out later, in Section \ref{sec:1/2},
that these two features
are not specific to
light spin-0 Dark Matter particles, but may apply as well to
spin-$\frac{1}{2}\,$ particles.

\section{Spin-0 Dark Matter annihilations through {$\,U$} exchanges}
\label{sec:spin0}

\subsection{Spin-0 annihilation cross sections}

\label{subsec:spin01}
\vspace{-1mm}
In the local limit approximation (valid,
in the annihilation case, for $\,2\,E \ll m_U$),
Dark Matter interactions may be described
by an effective Lagrangian density
involving the product of the Dark Matter ($\varphi$) and
quark and lepton ($f$) contributions
to the $U$ current, i.e.

\be
\label{effphi}
{\cal L}\ \ =\ \ \frac{C_U}{m_U^{\,2}}\ \ \
\varphi^*\ i \stackrel{\leftrightarrow}{\partial_\mu}\!\varphi\ \
(\,f_V\ \bar f\,\gamma^\mu f\ +\ f_A\ \bar f\,\gamma^\mu\gamma_5 f\,)\ \ ,
\ee
with ``Fermilike'' coupling constants
$\,G_V\approx C_U\,f_V/m_U^{\,2},$
$G_A\approx C_U\,f_A/m_U^{\,2}\,$.
$\,C_U$ and $f_V$ and/or $\,f_A$ denote the couplings
of the new gauge boson $U$
to the \hbox{spin-0} Dark Matter field  $\,\varphi\,$
and the matter fermion field $f$ considered, respectively.
(If the local limit approximation is not valid,
$\,U$ propagator effects may be taken into account by
replacing $\,-\,m_U^{\,2}\,$ by $\,s-m_U^{\,2}= 4\,E^2-m_U^{\,2}\,.$)
\,The contributions of the vector
and axial $f$ currents do not interfer for unpolarized cross sections
(an interference term would have to involve
the totally antisymmetric $\epsilon$ tensor and must therefore vanish),
and may be considered separately.
The $\,f_V$ coupling is invariant under Charge Conjugation,
while $\,f_A\,$ (which is likely to be absent,
otherwise we would generally have an unwanted axionlike behavior
of the new light gauge boson $U$ \footnote{In the presence
of an axial contribution ($f_A$) to the $U$ matter current,
the longitudinal polarization state of the $U$
would behave very much as an (unwanted) spin-0 axionlike particle, having
pseudoscalar couplings to quarks and leptons $\,\approx\,f_A\,m_{q,l}/m_U$
\cite{pfu}.})
is $\,C$-violating.

\vspace{3mm}

\noindent
\underline{No $S$-wave annihilation for spin-0 Dark Matter particles}

\vspace{2mm}
The threshold behavior of the annihilation cross section
$\,\sigma_{ann}\,(\varphi\bar\varphi\! \to f\bar f)\,$
may be understood easily from simple arguments based on Charge Conjugation.
The initial $\,\varphi\,\bar\varphi\,$ state
has $\,C=(-)^L = +\,$ in an $S$ wave ($L=0$).
The final $\,f \bar f\,$ state then also has $\,C'=(-)^{(L'+S')} = +\ $
(since angular momentum conservation requires $\,J'=J=0\,$).

\vspace{1mm}
In the case of a {\it axial\,} coupling \,($f_A$)\, to the fermion
field $f$, the relevant terms in the Lagrangian density (\ref{effphi}),
being $\,C$-violating, cannot induce the decay
$\,\varphi\bar \varphi_{\,S\hbox{\footnotesize -wave}}\to f\bar f\,$.

\vspace{1mm}

In the case of a {\it vector\,} coupling \,($f_V$),\,
the relevant terms in (\ref{effphi}) are indeed $\,C$-conserving,
but the $\,f\bar f\,$ final state, being vectorially produced
through the virtual production of a $U$ boson
(as if it were through a one-photon exchange),
must have $\,C'=-$
(while $\,C'=+\,$ from angular momentum conservation).

\vspace{1mm}
In both cases of vector and axial couplings
(or for a linear combination of them), there can be
no $S$-wave term in the annihilation cross section.
The dominant ($P$-wave) terms in $\,\sigma_{ann}\,v_{rel}\,$
are then proportional to the square of the Dark Matter particle velocity
in the initial state, i.e.
\be
\sigma_{ann}\,v_{rel}\,(\,\varphi\bar\varphi \to f\bar f\,) \,\propto\,
v_{dm}^{\,2}
\ee
at threshold.

\vspace{1mm}
Stated in other terms, the total $U$ charge-density
and current of a $\,\varphi\bar\varphi\,$ pair must vanish
at threshold. The annihilation amplitudes,
proportional to $\,i\,(p_1^{\,\mu}-\,p_2^{\,\mu})\,$,
\,vanish proportionally to the rest-frame momenta
$\,p_{dm}\,$ of the initial particles, or to $\,v_{dm}$,
\,as a result of the derivative nature of the $U$ coupling to scalar particles.

\vspace{1mm}

Let us now evaluate explicitly these annihilation cross sections
(as given in \cite{bf} at threshold, for $\,E\,\simeq\,m_{dm}\,$).
Still another way to obtain them, without calculation, from the corresponding
production cross sections in $\,e^+e^-\,$ annihilations,
will be given in Section {\ref{sec:relating}.

\vspace{3mm}

\noindent
\underline {Vector coupling to fermions:}

\vspace{2mm}
The following factor in the ``squared amplitude''
is easily evaluated, in the center of mass reference frame:

\vspace{-3mm}
\small
\be
\ba{ccl}
\!\!-\ (p_1-p_2)_\mu\,(p_1-p_2)_\nu\ \ \hbox{Tr} \
[\,(p\sl_3+m_f)\,\gamma^\mu\,(-p\sl_4+m_f)\,\gamma^\nu\,]
\hspace{-8cm}&&
\vspace{2.5mm}\\
\hspace{-8cm}&=&-
\,4\ (p_1-p_2)_\mu\,(p_1-p_2)_\nu\ \
\vspace{1.5mm}\\
&& \hspace{1cm}\left(\ -\,p_3^{\,\mu}\,p_4^{\,\nu}\,-\,p_3^{\,\nu}\,p_4^{\,\mu}\,+\,
g^{\mu\nu}\ (p_3.p_4+m_f^2)\,\right)
\vspace{2.5mm}\\
&=& -\ 16\ p_{1i}\,p_{1j}\ (\,2\,p_3^{\,i}\,p_3^{\,j}+2\,g^{ij}\,E^{\,2}\,)
\vspace{3mm}\\
&=&
\ \ \ \ 32\ \vec p_{dm}^{\ 2}\ E^2\ (1-\beta_f^{\,2}\,\cos^2\theta)
\ \ .
\ea
\ee
\normalsize

\noindent
Averaging over angles, and multiplying by
$\,C_U^{\,2}\,f_V^{\,2}\,/\,$
\linebreak
$(m_U^{\,2}-4\,E^2)^2,$ and by
$ \frac{1}{(2\,\pi)^2}\,\frac{1}{(2\,E)^4}$
{$\ 4\,\pi\ p_f$}\,$\frac{E_f}{2}=
\frac{1}{32\,\pi}\,\frac{1}{E^2}\ \beta_f\,$
for the phase space integration, we get
\be
\label{sigmaannsc}
\ba{c}
\displaystyle
\sigma_{ann}\,v_{rel}\ =\
\frac{2}{3\,\pi}\ \,v_{dm}^{\,2}\
\hbox{\small $\displaystyle \frac{C_U^{\,2}\,f_V^{\,2}}{(m_U^{\,2}-4\,E^2)^2}$}\
E^2\
{\hbox{\Large(}}\,\frac{3}{2}\,\beta_f-\frac{1}{2}\,\beta_f^3\,{\hbox{\Large)}}\
\vspace{4mm}\\
\displaystyle
\ \ =\ \frac{2}{3\,\pi}\ \,v_{dm}^{\ 2}\
\frac{C_U^{\,2}\,f_V^{\,2}}{(m_U^{\,2}-4\,E^2)^2}\
\hbox{\footnotesize$\displaystyle \sqrt{\ \hbox{\normalsize 1}-\frac{m_f^{\,2}}{E^2}}$}\ \,
{\hbox{\LARGE(}}\,E^2+\,\frac{m_f^{\,2}}{2}\,{\hbox{\LARGE)}}\,.
\ea
\ee

\vspace{2mm}
\noindent
This reduces, at threshold
\,($\,s=4\,E^2\simeq 4\ m_{dm}^{\,2}\,$),
\,to the same expression as in \cite{bf} (choosing $\,f_{U_l}=f_{U_r}=f_V\,$).
We recognize,
in terms of the velocity parameter
$\,\beta_f = v_f\ (/c)=\hbox{$(\ 1-m_f^{\,2}/E^2)^{1/2}$}$,
\,the usual kinematic factor relative to the vectorial production
of a pair of spin-$\frac{1}{2}$ Dirac fermions,
\be
\label{phasespacevector}
\frac{3}{2}\ \beta_f\ -\ \frac{1}{2}\ \beta_f^{\,3}\ \ =\ \
\hbox{\footnotesize $ \displaystyle
\sqrt{\ \hbox{\normalsize 1}-\frac{m_f^{\,2}}{E^2}}$}\ \ \,
{\hbox{\Large(}}\,E^2+\,\frac{m_f^{\,2}}{2}\,{\hbox{\Large)}}\ \ .
\ee

\vspace{3mm}

\noindent
\underline{Axial coupling to fermions:}

\vspace{2mm}
We can still use the previous calculation,
replacing $f_V$ by $f_A$,
and changing $\,m_f^{\,2}\,$ into $\,-\,m_f^{\,2}\,$
within the expression of the ``squared amplitude'' $|{\cal A}|^2\,$.
$\,p_3.p_4+m_f^{\,2}$ $=2\,E_f^{\,2}\,$ is replaced by
$\,p_3.p_4-m_f^{\,2}\,= \,2\,p_f^{\,2}\,$, \,and
$\ (1-\beta_f^{\,2}\,\cos^2\theta)\ $ by
$\ (\beta_f^{\,2}-\beta_f^{\,2}\,\cos^2\theta)\,=\, \beta_f^{\,2}\,\sin^2\theta\,$.
The kinematic factor (\ref{phasespacevector})
which appears for the vectorial production
of the $\,f\bar f\,$ pair gets simply replaced by the corresponding factor
$\,\beta_f^{\,3}\,$ appropriate to the axial production
of \hbox{spin-$\frac{1}{2}$} particles.
Expressing the latter in terms of $\,m_f\,$ and the
Dark Matter particle energy $\,E_{dm}=E_f=E\,$, \,we get
\be
\ba{c}
\label{annaxial}
\displaystyle
\!\!\sigma_{ann}\,v_{rel}\ = \ \frac{2}{3\,\pi}\ \,v_{dm}^{\ 2}\
\frac{C_U^{\,2}\,f_A^{\,2}}{(m_U^{\,2}-4\,E^2)^2}\ E^2\
{\hbox{\LARGE(}}\,\,1-\frac{m_f^{\,2}}{E^2}\,{\hbox{\LARGE)}}^{3/2}.
\ea
\ee
Again this reduces, at threshold ($\,E\simeq m_{dm}\,$),
\,to the same expression as obtained from \cite{bf} (choosing
$\,f_{U_l}=-\,f_{U_r}=f_A$).

\vspace{2mm}
If the $U$ coupling to the fermion field $f$ includes
both vector and axial contributions,
the annihilation cross section is the sum of the two
contributions (\ref{sigmaannsc}) and (\ref{annaxial}).

\subsection{Constraints on the $U$ couplings}

\vspace{-1mm}
Numerically, and to get an idea of the size of the couplings
(depending also on the masses $\,m_U$ and $\,m_{dm}$) \,required
to get appropriate values of the annihilation cross sections at decoupling
(i.e. about \,8 \,to 10 picobarns, \,cf.~\,Section \ref{sec:relic}), we can write
the above expressions (\ref{sigmaannsc},\ref{annaxial}) as
\be
\label{sigmanum}
\vspace{2mm}
\sigma_{ann} v_{rel}\ \simeq\
\hbox{\small $\displaystyle
\frac{v_{dm}^{\,2}}{.16}
\
\left(\frac{C_U\,f_{V,A}}{10^{-6}}\right)^2
\
\left(\frac{m_{dm}\times 3.6\ \rm{MeV}}{m_U^{\,2}-4\,m_{dm}^{\,2}}\right)^2
$}
\,\hbox{pb}\,,
\ee
still to be multiplied by the appropriate kinematic
factor ($<1$) \,relative to the vectorial ($\,\frac{3}{2}\,\beta_f-\frac{1}{2}\,\beta_f^{\,3}$)
\,or axial ($\beta_f^{\,3}$) \,production of spin-$\frac{1}{2}$
particles\ \footnote{$v_{dm}^{\,2}\simeq .16\ c^2$ is associated with
a kinetic energy $\,\simeq .09$ $m_{dm}$, \,which would correspond roughly, to fix ideas, to
$\,\frac{3}{2}\ T_F=\frac{3}{2\,x_F}\ m_{dm}$ with $\,x_F\simeq 16$ or 17.}.
We shall in fact consider mostly vectorial couplings of the $\,U$ to ordinary matter
($f_{V}$), with values much smaller than the electric charge \hbox{($e\simeq .3$)}
\,by several orders of magnitude.
The resulting $\,U$ boson effects on ordinary particle physics processes,
charged lepton $g-2$, \,etc., then appear sufficiently small \cite{bf}.

\vspace{2mm}

In particular, for a vectorially coupled $U$ boson somewhat heavier than the electron
but lighter than the muon, the comparison between the additional $U$ contributions
to the muon and electron anomalies and the possible difference between the experimental and Standard Model values
indicates that \cite{bf,g-2}
\be
\ba{ccccc}
\delta a_\mu &\simeq &
\displaystyle
\frac{f_{V\mu}^{\,2}}{8\,\pi^2} &\simeq& (\,2\pm2\,)\ 10^{-9}
\ \ ,
\vspace{2mm}\\
\delta a_e &\simeq & \displaystyle
\frac{f_{Ve}^{\,2}}{12\,\pi^2}\ \frac{m_e^{\,2}}{m_U^{\,2}}
&\simeq& \,(\,4\pm 3\,)\ 10^{-11}\ \ ,
\ea
\ee
so that
\be
f_{V\mu}\simle \,6\ 10^{-4}\ \ ,\ \ \ \ f_{Ve}\simle \,2\ 10^{-4}\ m_U \hbox{(MeV)}\ \ .
\ee
One should also have, for a $U$ mass larger than a few MeV's,
\be
|f_{V\nu}\,f_{Ve}|\ \simle\ G_F\ m_U^{\,2}\ \simeq\ 10^{-11}\ (m_U(\hbox{MeV}))^2\ \
\ee
so that $U$ exchanges do not modify excessively neutrino-electron low-energy elastic scattering
cross sections, in good agreement with Standard Model values \cite{nu}.
This requires that the $U$ couplings to neutrinos, at least, be
sufficiently small. This also requires, conversely, that the $U$ coupling $c_U$ to Dark Matter
be not too small, so as to get, from eq.\,(\ref{sigmanum}),
appropriate values of the annihilation cross section
$\,<\sigma_{ann} v_{rel}>\,$
\,\footnote{If, in addition, all $f_V$'s are of the same order,
they should then be smaller than about $\,.3\ 10^{-5}\,
m_U$(MeV). Expression (\ref{sigmanum}) of
$\,<\!\sigma_{ann}v_{rel}\!>\,$  then implies
that the $U$ coupling to Dark Matter $c_U$ should be {\it \,larger\,} 
 than about $\,.2\ (m_U^{\,2}-4\,m_{dm}^{\,2})/m_U\,m_{dm}$ (for $\,<\!\sigma_{ann}v_{rel}\!>\,$
to be larger than 4 or 5 pb),
in which case  the self-interactions of Dark Matter would be significant,
much stronger than ordinary weak interactions, by several orders of magnitude
(depending also on the energy considered)!
These particles would continue interacting with themselves for some time after the freeze out of their annihilation reactions. They would also still interact with ordinary particles,
subsequently leading to an enhancement of the free-streaming scale\,\cite{damp},
as a result of these interactions.}.
\vspace{2mm}

The same cross section formulas (with vector couplings $f_V$), etc.,
may also be used, as we shall see in Section
\ref{sec:1/2}, in the case of spin-$\frac{1}{2}$ Dark Matter particles, 
as well as for spin-0 particles.

\section{Relating production and annihilation cross sections}
\label{sec:relating}

The production and annihilation cross sections
of Dark Matter particles may be easily related, as the corresponding amplitudes
are related by $CPT$ (or simply $T$, when $\,CP\,$ invariance holds).
When computing cross sections $\,\sigma\, v_{rel}\,$
we sum on final state polarisations while averaging over initial ones.
The integration of the squared modulus of the transition amplitudes
($\,|\,{\cal A}\,|^2$)
\,-- identical when one appropriately exchanges
the initial and final states --\,
over the final particle momenta,
makes the velocity of the latter particles
($\,\beta_f = v_f\,$,
or $\,\beta_{dm}=v_{dm}$, \,for $\,f,\bar f\,$
or Dark Matter particles) appear.
The production (in $\,e^+e^-$ scatterings) and annihilation
cross sections are then related by:

\vspace{-5mm}
\be
\label{prodann}
\hbox{\small$\displaystyle
\sigma_{prod}\,v_{e}\,(e^+e^-\!\to\varphi\,\bar\varphi)\,/v_{dm}
\ \equiv\
\frac{1}{4}\
\sigma_{ann}\,v_{dm}\,(\varphi\,\bar\varphi \to e^+e^-)\,/v_e\,.
$}
\ee

\vspace{1mm}

From the usual electromagnetic pair production cross section
of charged \hbox{spin-0} particles in $\,e^+e^-\,$ annihilations
(neglecting $\,m_e$),
\be
\sigma_{prod}^{\ (\gamma)}\ \ =\ \ \frac{4\,\pi\,\alpha^2}{3\,s}\ \
\frac{1}{4}\ \,\beta_{dm}^{\,3}\ =\
\frac{e^4}{48\ \pi\ s}\ \,\beta_{dm}^{\,3}\ \ ,
\ee
we immediately get (replacing $\,e^4/s^2\,$ by
$\,C_U^{\,2}\,f_{\,V}^2\,/$ \linebreak
$(s-m_U^{\,2})^2\ $)
\,the production cross section, through $U$ exchanges,
of neutral spin-0 Dark Matter particles,
\be
\label{sigmaprod}
\sigma_{prod}\,(e^+e^-\!\to\varphi\,\bar\varphi)\ =\
\frac{1}{12\,\pi}\ \frac{C_U^{\,2}\,f_{\,V}^2}{(4\,E^2-m_U^{\,2})^2}\
 \ E^{2}\
\beta_{dm}^{\,3}\ \,.
\ee

\vspace{3mm}
Multiplying it by $\,v_{rel}\simeq 2$,
\,by the spin factor \,4\, and the velocity ratio
$\,(\beta_e\simeq 1)/(\beta_{dm}=v_{dm})\,$ appearing in (\ref{prodann}),
we get the corresponding annihilation cross section,
\be
\label{ann}
\sigma_{ann}\,v_{rel}\, (\,\varphi\,\bar\varphi\to\,e^+e^-\,)\ =\
\frac{2}{3\,\pi}\ v_{dm}^{\,2}\
\frac{C_U^{\,2}\,f_{\,V}^2}{(m_U^{\,2}-4\,E^2)^2}\ \ E^2\ ,
\ee
which, once the kinematic factor
$\,\frac{3}{2}\,\beta_f-\frac{1}{2}\,\beta_f^{\,3}$
\,(taking into account the effect of non-vanishing $\,m_e$)
\,is reintroduced, coincides precisely with eq.\,(\ref{sigmaannsc}).
In a similar way replacing $\,f_V$ by $\,f_A\,$, and reintroducing the
appropriate kinematic factor $\,\beta_f^{\,3}$,
\,we recover eq.\,(\ref{annaxial})
for the annihilation cross section $\,\sigma_{ann}\,v_{rel}\,$
through an axial coupling to the matter fermion field $f$.

\vspace{2.5mm}
The $\,v_{dm}^{\,2}\,$ suppression factor in the annihilation cross section
$\,\sigma_{ann}\,v_{rel}\,$ of spin-0 particles $\,\varphi\,$
appears simply as {\it a reflection by $\,CPT\,$ of the
well-known $\,\beta^3\,$ factor\,}
for the pair production of spin-0 particles in $\,e^+e^-$ annihilations
(with, in both cases, a $P$ wave for the $\,\varphi\,\bar\varphi\,$ state).

\vspace{2.5mm}

Equation (\ref{sigmaprod}) may be used to discuss
the pair production of spin-0 Dark Matter particles
in $\,e^+e^-$ annihilations.
When the $U$ and $\,\varphi\,$ particles are light
it may be written under any of the equivalent forms:
\be
\ba{ccl}
&&\hspace{-1cm}
\sigma_{prod}\,(\,e^+e^-\to\varphi\,\bar\varphi\,) \ \simeq\
\hbox{\small $\displaystyle
\frac{C_U^{\,2}\,f_{\,V}^2}{48\,\pi\ s}\ \simeq\
\frac{C_U^{\,2}\,f_{\,V}^2}{192\,\pi\ E_e^{\,2}}\
$}
\vspace{3mm} \\
&\simeq& \ \ \displaystyle{\frac{\alpha_U\,\alpha_V}{\alpha^2}\ \
\frac{21.7\ \hbox{nb}}{s\,(\hbox{\footnotesize GeV}^2)}
}\
\simeq\ \ \displaystyle{ C_U^{\,2}\,f_{V}^{\,2}\ \
\frac{2.58\ \mu\rm{b}}{s\,(\hbox{\footnotesize GeV}^2)}
}
\vspace{3mm} \\
&\simeq&   \displaystyle{ \left(\frac{C_U\,f_V}{10^{-6}}\right)^2\ \
\frac{2.6\ 10^{-42}\ \hbox{cm}^2}
{\left(\sqrt s\,(\hbox{\footnotesize GeV})\right)^2}\ \ .
}
\ea
\ee

For the relevant values
of $\,C_U f_V$ \,(or conceivably
$\,C_U f_A$) considered,
these production cross sections get very small at high energies, much below
neutrino production cross sections,
so that the direct production of such Dark Matter particles
is in general not expected to lead to easily observable signals in
$e^+e^-$ annihilations \cite{bf}.

\vspace{2mm}
Let us now turn to spin-$\frac{1}{2}$ particles.
Their pair production through an axial coupling involves
a $\,\beta_{dm}^{\,3}\,$ factor
\,\footnote{As, also, for the decay
of a massive spin-1 gauge boson into two ``inos'',
or the pair production \,(neglecting $\,m_e$) \,of photinos in $e^+e^-$
annihilations, through (effective) axial couplings.}.
It reflects precisely,
as we saw, in a $\,v_{dm}^{\,2}\,$ suppression factor
for the corresponding annihilation cross section at threshold
(at least as long as the masses $m_f$ of the produced fermions
are neglected, a point to which we shall return
in Section \ref{sec:1/2}).

\vspace{1mm}

Since the {\it \,production\,} cross sections
of spin-0 and \hbox{spin-$\frac{1}{2}$} particles in $\,e^+e^-\,$ annihilations
are given by similar formulas
we expect that the corresponding {\it \,annihilation\,} cross sections
into $\,f\bar f\,$ pairs be given, also,
by similar formulas.
Still it is essential to clarify
under which circumstances
these annihilation cross sections will continue to behave at threshold
like $\,v_{dm}^{\,2}$, \,when non-vanishing fermion masses $\,m_f\,$
are taken into account.

\section{Spin-$\frac{1}{2}$ Dark Matter particles}
\label{sec:1/2}

Again the vector and axial couplings of the $\,U\,$ boson
to the fermions $\,f\,$ may be considered independently.
The effective Lagrangian density
(similar to the one responsible
for the effective interactions of photinos with matter
through $\,\tilde q\,$ or $\,\tilde l\,$ exchanges)
may now be written as
\be
\label{effchi}
{\cal L}\ \ =\ \ \frac{C_U}{2\ m_U^{\,2}}\ \ \
\bar\chi\,\gamma_\mu\gamma_5\,\chi\ \
(\,f_V\ \bar f\,\gamma^\mu f\ +\ f_A\ \bar f\,\gamma^\mu\gamma_5 f\,)\ \ .
\ee
In contrast to the previous case of a spin-0 field
$\,\varphi\,$,
the coupling $\,f_V\,$ is now $C$-violating while $\,f_A\,$
(still normally presumed to be absent
as it would be related with an unwanted axionlike behavior of the $U$ boson)
is $C$-conserving.

\subsection{$\,\sigma_{ann}\,$ vanishes at threshold, for vector couplings
\ \ \ \ \ \ \ \ \ \ \ \ --\ \  but not for axial ones}

\vspace*{-1mm}

At threshold the antisymmetry of our 2-Majorana $\,\chi\,\chi$ state
(in an $S$ wave) imposes that the total spin be $J=S=0$,
so that the production (indifferently through vector and/or axial couplings)
of massless fermion pairs $\,f\bar f\,$
with total angular momentum $\,\lambda=\pm 1\,$ along their direction
of propagation is forbidden.
But the ($S$-wave) annihilation cross section $\,\sigma_{ann}\,v_{rel}\,$
could in principle include, at threshold, non-vanishing contributions
proportional to $\,m_f^{\,2}$.

\vspace{1mm}

Our initial 2-Majorana $\,\chi\,\chi$ state has $\,C=+\,$.
\,If it is in an $S$ wave (so that $J=0$), the
final $\,f\bar f\,$ state must have $\,C'=(-)^{(L'+S')}=+$
\,(with $\,L'=S'$ from angular mo\-men\-tum conservation).
It follows immediately that the ($C$-violating) vectorial coupling $\,f_V\,$
cannot contribute to the $S$-wave
$\,\varphi\bar\varphi\to f\bar f\,$ annihilation amplitude.

\vspace{1mm}
A second reason is that a $\,U\,$ boson with a vectorial coupling
to the $\,f\,$ can only produce a $\,f \bar f\,$ pair
with \hbox{$\,C'=-$,}
while it must have  $\,C'=+\,$ from angular momentum conservation.
For either reason, the $\,f_V\,$
contribution to the $S$-wave annihilation cross section must vanish,
so that
\be
\displaystyle
\sigma_{ann}\,v_{rel}\ (\,\chi\,\chi \to\bar f\,f\,)\ \ \propto
\hbox{\large$\displaystyle \ \ v_{dm}^{\,2}$}
\ \ \
\hbox{at threshold}\ \ .
\ee

\vspace{1mm}
The situation would clearly be different for the production
of massive fermion pairs $\,f\bar f\,$
through an {\it \,axial\,} current (with $C=+$),
rather than a {\it \,vector\,} current (with $C=-$).
The $\,f_A\,$ contribution to the $S$-wave annihilation
cross section has now no reason to vanish, as soon
$\,m_f\neq 0\,$, since:

i) the corresponding $\,C$-conserving operator in (\ref{effchi})
can indeed induce a $\,\chi\chi\to f\bar f\,$
transition from a $\,C=+\,$ to a $\,C'=+\,$ state;

ii) the axial fermionic current $\,\bar f\gamma_\mu\gamma_5\,f$,
\,being $C$-even,
is capable of creating a $\,f \bar f\,$ pair in a $\,C'=+\,$ state.

\vspace{1mm}

Constant terms proportional to $m_f^{\,2}\,$
(undesirable for us here, at least for light Dark Matter particles $\,\simle 100$ MeV)
\,do then appear in the annihilation cross section
$\,\sigma_{ann}\,v_{rel}\,$.

\vspace{2mm}

It is remarkable that the constraint of {\it \,vector\,} couplings
of the light $U$
to the matter fermions $f$, obtained here from the requirement
of annihilation cross
sections behaving like $\,v_{dm}^{\,2}\,$,
\,is essentially the same as already necessitated from the fact
that such a light $\,U\,$ boson (given the masses and couplings
considered), would have an unacceptable axionlike behavior
if it had sizeable axial couplings $\,f_A\,$ to the matter fermions $f$ \cite{pfu}.

\subsection{From spin-0  to spin-$\frac{1}{2}$ Dark Matter
annihilation cross sections}

\vspace*{-1mm}

We now intend to compare the pair {\it \,annihilation\,} cross section for
(Majorana) spin-$\frac{1}{2}$ particles $\chi$
(with axial coupling $\,\frac{C_U}{2}\,$), and
for spin-0 particles of $\,U$-charge $\,C_U,\,$
by relating them to the corresponding {\it \,production\,} cross sections
in $\,e^+e^-\,$ annihilations.

\vspace{2mm}

In the limit of vanishing $\,m_e\,$ the production cross section of a pair of
Dirac particles ($\,\psi\,\bar\psi\,$) through an axial coupling
($\,C_U\ \bar\psi\,\gamma_\mu\gamma_5\,\psi\,$) to the $U$
is proportional to $\,\beta_{dm}^{\,3}$. It is related
to the production cross section for a pair of
spin-0 particles \,($\,\varphi\bar\varphi\,$),
proportional to $\,\frac{1}{4}\,\beta_{dm}^{\,3}$
\,(with in both cases a $\,\beta_{dm}^{\,3}\,$ factor
associated with a $P$-wave production
of these particles in the final state), by:
\be
\label{prodprod}
\sigma_{prod} \,(\,e^+e^-\,\to\,\psi\,\bar\psi\,)\ \ \equiv\ \ 4\
\sigma_{prod} \,(\,\ e^+e^-\,\to\,\varphi\,\bar\varphi \,)\ \ .
\ee

\vspace{1mm}
By using the relations of Section \ref{sec:relating} between annihilation
and production cross sections (as expressed by
(\ref{prodann}) for spin-0 particles),
we get from eq.\,(\ref{prodprod}) the following relation between the (Dirac)
spin-$\frac{1}{2}$ and \hbox{spin-0} annihilation cross sections:
\be
\label{annpsiphi}
\sigma_{ann}\,(\,\psi\,\bar\psi\,\to\,e^+e^-\,)\ \ =\ \
\sigma_{ann}\,(\,\varphi\,\bar\varphi\,\to\,e^+e^-\,)\ \ .
\ee

\vspace{1mm}

To relate the annihilation cross sections of Dirac and Majorana particles
we can use the following trick:
by writing the decomposition
$\,\psi=(\chi-i\chi')/\sqrt 2\ $ of
the Dirac spinor field $\,\psi\,$,
so that
\be
\ba{c}
\bar\psi\,\gamma_\mu\gamma_5\,\psi \ \ =\ \
\frac{1}{2}\ \,\bar\chi\,\gamma_\mu\gamma_5\,\chi \ +\
\frac{1}{2}\ \,\bar\chi'\,\gamma_\mu\gamma_5\,\chi'
\ea
\ee
and considering an initial state
in which each of the two annihilating particles
is either  a $\,\psi\,$ or a $\,\bar\psi\,$ (i.e. just as well, equivalently,
either a $\,\chi\,$ or a $\,\chi'\,$),
we see that the pair annihilation cross section
of Dirac particles ($\,\psi$, with axial coupling $\,C_U$)
is the same as for Majorana particles ($\,\chi$, with axial coupling $\,C_U/2$).
It follows that:
\be
\label{annpsiphi2}
\ba{ccl}
\sigma_{ann}\,(\,\chi\,\chi\,\to \,e^+e^-\,)&=&
\sigma_{ann}\,(\,\psi\,\bar\psi\,\to\, e^+e^-\,)
\vspace{1.5mm} \\
&=&\sigma_{ann}\,(\,\varphi\,\bar\varphi\,\to\, e^+e^-\,)\ \ ,
\ea
\ee
the latter being given by eqs.~(\ref{sigmaannsc}) or \,(\ref{ann}).
Altogether we get:
\be
\label{annprime}
\sigma_{ann}\,v_{rel}\,(\,\chi \,\chi \to e^+e^-)
\ \ =\ \
\frac{2}{3\,\pi}\ v_{dm}^{\,2}\
\hbox{\small $\displaystyle \frac{C_U^{\,2}\,f_{\,V}^2}{(m_U^{\,2}-4\,E^2)^2}$}
\ \ E^2\,.
\ee

\vspace{1mm}

We can now take into account explicitly
the effect of the electron mass in the final state.
Our previous arguments
showed that no $\,S$-wave annihilation cross section may be induced
from a non-vanishing $m_e\,$, in the case of a vectorial coupling $\,f_V$
(in contrast with $\,f_A\,$).
The only expected effect of a non-vanishing $\,m_e$,
or more generally of fermion masses $\,m_f$,
\,will simply be a multiplication of
(\ref{annprime}) by the usual kinematic factor for
the vectorial production of a $\,f\bar f\,$ pair,
$\,\frac{3}{2}\ \beta_f\,-\,\frac{1}{2}\ \beta_f^{\,3}\,$.
This yields:
\be
\label{sigmaann12}
\ba{l}
\sigma_{ann}\,v_{rel}\,(\,\chi \,\chi \to e^+e^-) \hspace{5cm}
\vspace{2mm} \\
\ \ =\ \frac{2}{3\,\pi}\ \,v_{dm}^{\,2}\ \
\hbox{\small$ \displaystyle
\frac{C_U^{\,2}\,f_V^{\,2}}{(m_U^{\,2}\,-\,4\,E^2)^2}
$}
\ \
E^2\ (\,\frac{3}{2}\,\beta_f-\frac{1}{2}\,\beta_f^3\,)\
\vspace{3mm}\\
\displaystyle
\ \ =\ \frac{2}{3\,\pi}\ \,v_{dm}^{\ 2}\
\frac{C_U^{\,2}\,f_V^{\,2}}{(m_U^{\,2}-4\,E^2)^2}\
\hbox{\small$\displaystyle \sqrt{\ 1-\frac{m_f^{\,2}}{E^2}}$}\ \,
{\hbox{\Large(}}\,E^2+\,\frac{m_f^{\,2}}{2}\,{\hbox{\Large)}}\,.
\ea
\ee

\vspace{2mm}
Remarkably enough, this cross section is actually identical to the cross section
(\ref{sigmaannsc})
for the pair annihilation of spin-0 Dark Matter candidates\,!
In particular, we get in both cases the same
$\,v_{dm}^{\,2}\,$ suppression factor of the annihilation cross sections,
as desirable to avoid an excessive production of gamma rays
originating from residual light Dark Matter annihilations.
The (collisional and free-streaming) damping effects  \cite{bf,damp} associated
with such particles are also, in both cases, sufficiently small.

\vspace{1mm}

The effect of $\,m_f\,$ in the case of an axial coupling to fermions,
however, will now be much more drastic than a simple multiplication by $\,\beta_f^{\,3}$,
\,since new terms not behaving like
$\,v_{dm}^{\,2}\,$
(and proportional to $\,m_f^{\,2}\,$) will appear
in the annihilation cross section.

\subsection{Direct evaluation of spin-$\bf \frac{1}{2}$ cross sections}

\vspace*{-2mm}

Now that we know the result, at least in the case of a vector coupling $\,f_V$,
without performing any explicit calculation,
we can verify it explicitly.
Since the annihilation cross section for a pair of
{\it \,Majorana\,} particles ($\,\chi$, with an axial coupling
$\,C_U\,\frac{1}{2}\,\gamma^\mu\gamma_5\,$ to the $U$ boson),
is the same as for {\it \,Dirac\,} fermions
($\,\psi$, with an axial coupling $\,C_U\,\gamma^\mu\gamma_5\,$ to the $U$),
we can evaluate the ``squared amplitudes'',
and resulting cross sections,
as if we were dealing with such Dirac fermions.
With an overall  $\,\frac{1}{4}\,$ factor
from the average on the incoming spin states,
a first trace factor corresponding to the pair
annihilation of Dark Matter particles
through an axial coupling,
and a second one to the $\,f\bar f\,$ pair production
through a vector coupling, we evaluate:

\vspace{4mm}
\bea
\label{calc}
\hbox{\small$\displaystyle
\ba{l}
\vspace{-14mm} \\
\frac{1}{4}\ \,\hbox{Tr} \,
[(p\sl_1+m_{dm})\,\gamma_\mu\gamma_5\,(-p\sl_2+m_{dm})\,\gamma_\nu\,\gamma_5]\
\vspace{2mm}\\
\hspace{2cm}\hbox{Tr} \
[(p\sl_3+m_f)\,\gamma^\mu\,(-p\sl_4+m_f)\,\gamma^\nu]\ \ \ \ \
\vspace{2mm}\\
=\ 4\ \,\left(\,- \,p_{1\,\mu}\,p_{2\,\nu}\,-\,p_{1\,\nu}\,p_{2\,\mu}\,+\,
g_{\mu\nu}\,(p_1.p_2\,-\,m_{dm}^2)\,\right) \hspace{3cm} \vspace{2mm}\\
\hspace{2cm}
\left(\ -\,p_3^{\,\mu}\,p_4^{\,\nu}\,-\,p_3^{\,\nu}\,p_4^{\,\mu}\,+\,
g^{\mu\nu}\,(p_3.p_4\,+\,m_f^2)\,\right)\ \ \vspace{2mm}\\
=\ 4\ \,\left[\ 2\ p_1.p_3\ p_2.p_4\ +\ 2\ p_1.p_4\ p_2.p_3\ - \ 2\ p_3.p_4\ m_{dm}^2\
\right. \vspace{2mm}\\
\hspace{2cm} \left.
+ \ 2\ p_1.p_2\ m_f^2\ -\ 4\ m_{dm}^2\,m_f^2\ \right]\ \
\vspace{3mm}\\
=\ 4\ \,\left[\ 2\,(E^2-p_{dm}\,p_f\,\cos\theta)^2\ +\
2\,(E^2+p_{dm}\,p_f\,\cos\theta)^2\ \right.
\vspace{2mm}\\
\hspace{4mm}
\left. -\,2\,(2\,p_f^{\,2}+m_f^{\,2})\, m_{dm}^{\,2}
+2\,(2\,p_{dm}^{\,2}\!+m_{dm}^{\,2})\, m_f^{\,2}
-4\, m_{dm}^{\,2}\,m_f^{\,2}\right]
\vspace{4mm}\\
\hbox{\small $
=\ \ 4\ \left[\ 4\ p_{dm}^{\,2}\ p_f^{\,2}\ (\,1+\,\cos^2\theta\,)
\ +\  8\ p_{dm}^{\,2}\,m_f^{\,2}\ \right]\
$}
\vspace{-2mm}\\
\ea
$}
\nonumber
\eea
\be
\hbox{\small $\displaystyle
\ba{ccc}
\hspace{1cm} \to\ \ \ \hbox{\small $
16\ \ \beta_{dm}^{\,2}\ E^2\
\,\left(\,\frac{4}{3}\ \,p_f^{\,2}\,+\,2\ m_f^{\,2}\,\right)\ \ \
$}
\vspace{3mm}\\
\hspace{1cm} \ =\ \ \ \hbox{\small $
32\ \ \frac{2}{3}\ \,v_{dm}^{\,2}\ \ E^2\ \
\,\left(\,E^2\,+\,\frac{m_f^{\,2}}{2}\,\right)\ \ .
$}
\ea
$}
\ee

\vspace{2mm}
\noindent
Averaging over angles as done above, reintroducing the factor
$\,C_U^{\,2}\,f_V^{\,2}/(m_U^{\,2}-4\,E^2)^2$,
\,and multiplying (as for a spin-0 particle in Subsection \ref{subsec:spin01}) by
$\,\frac{1}{32\,\pi}\,\frac{1}{E^2}\ \beta_f\,$ for the phase space integration,
\,we recover precisely the previous expression (\ref{sigmaann12}) for
the annihilation cross section of \hbox{spin-$\frac{1}{2}$} Dark Matter particles,
a result also identical to the one obtained in (\ref{sigmaannsc})
for spin-0 particle annihilations.

\vspace{4mm}

Let us also give for completeness the corresponding spin-$\frac{1}{2}$
annihilation
cross section in the case of an axial matter fermion coupling $\,f_A\,$.
By changing $\,m_f^{\,2}\to-\,m_f^{\,2}\,$
at appropriate places in the calculation of eq.\,(\ref{calc}),
this expression of the squared amplitudes gets replaced by the now symmetric one

\vspace{4mm}
\be
\label{calc2}
\ba{l}
\vspace{-14mm} \\
\hbox{\small$\displaystyle
4\ \,\left[\ 2\ p_1.p_3\ p_2.p_4\ +\ 2\ p_1.p_4\ p_2.p_3\ - \ 2\ p_3.p_4\ m_{dm}^2\
\right.
$}
 \vspace{2mm}\\
\hbox{\small$\displaystyle
\hspace{4cm} \left.
- \ 2\ p_1.p_2\ m_f^2\ +\ 4\ m_{dm}^2\,m_f^2\ \right]\ \
$}
\vspace{3mm}\\
=\ \,16\ [\ \,p_{dm}^{\,2}\,p_f^{\,2}\  (1+\cos^2\theta)
\,+\,m_{dm}^{\,2}\,m_f^{\,2}\ \,]\ \ ,
\ea
\ee
which leads to:
\be
\ba{ccl}
\label{sigmaannsc3}
\displaystyle
\!\!\!\!\sigma_{ann}\,v_{rel}&=&
\hbox{\small$\displaystyle \frac{1}{2\,\pi}\ \,
\frac{C_U^{\,2}\,f_{A}^2 }{(m_U^{\,2}\,-\,4\,E^2)^2}$}\ \
\hbox{$\sqrt{\,1-\frac{m_f^2}{E^2}}$}\ \
\vspace{3mm} \\
&& \hspace{3mm}
\left[\ \,\frac{4}{3}\ \,(E^2-m_f^2)\ v_{dm}^{\,2}  \,+\,
\hbox{\large $ \frac{m_{dm}^{\,2}}{E^2}  $}
\ m_f^2\ \right]\,.
\ea
\ee

\noindent
It does coincide with (\ref{sigmaann12})
(replacing $\,f_A\,$ by $\,f_V$) in the limit of vanishing fermion masses $\,m_f$,
for which there is no physical distinction between vector and axial matter fermion couplings,
so that we get in both cases a $\,v_{dm}^{\,2}\,$ suppression factor
in the annihilation cross section.
But, as anticipated earlier, this overall $\,v_{dm}^{\,2}\,$ factor
no longer subsists for non-vanishing fermion masses $m_f$,
for which one recovers a non-vanishing $S$-wave term
in the annihilation cross section (\ref{sigmaannsc3}), proportional to $\,m_f^{\,2}$.

\subsection{Final remarks}

Altogether, in the case of a vector coupling $f_V$ of the $U$ boson
to quark and lepton fields $f$,
spin-$\frac{1}{2}$ Dark Matter particles have
the required characteristics for
Light Dark Matter (LDM) particles annihilating into $\,e^+e^-\,$ pairs,
just as well as spin-0 particles.
In both cases, $U$-induced Dark-Matter/electron interactions
should be significantly stronger than ordinary weak interactions
at low energy (but weaker at high energies),
which requires the $U$ to be more strongly coupled to Dark Matter
than to ordinary matter
\,-- also resulting in significant $\,U$-induced Dark Matter self-interactions.
Finally, light spin-$\frac{1}{2}$ Dark Matter particles
appear more attractive than \hbox{spin-0} ones,
as the smallness of their mass is easier to understand, and
provide valuable alternative scenarios to be discussed and confronted with the standard ones.

\vspace{2mm}

\appendix

\section{\ {E\lowercase {valuating {\bf $\, \boldmath x$}}}$_F$}
\label{app:xf}

For a Dark Matter particle of mass $m_{dm}$ with $g$ degrees of freedom (including antiparticles),
freezing out in the non-relativistic regime at a temperature $\,T_F= m_{dm}/x_F$,
the residual number density
(evaluated as if in equilibrium) at $T_F$ is
\be
N_{dm}\ \simeq\ g\ \left(\frac{m \,T_F}{2\pi}\right)^{3/2}\,\displaystyle e^{-\frac{m}{T_F}}\ \,
\simeq\, \ g\ \, T_F^{\,3}\ \left(\frac{x_F}{2\pi}\right)^{3/2}\displaystyle e^{-x_F}\ .
\ee
We work here in the na\"{\i}ve approximation in which the residual abundance of Dark Matter
particles is taken to be determined from its equilibrium value at the freeze out temperature $T_F$.
If this occurs after $e^+e^-$ annihilations, Dark Matter particles get diluted by the expansion of the Universe, as for photons, so that their present density reads:
\be
\label{A2}
N_{\circ\,dm}\ \ =\ \ g\ \,T_{\circ\,\gamma}^{\ \,3}\ \left(\frac{x_F}{2\pi}\right)^{3/2}
 \displaystyle e^{-x_F}\ \ ,
\ee
with $\,T_{\circ\,\gamma}^{\,\ }\simeq 1685$ cm$^{-3}$.
For particles in the $\,\approx 1$ MeV mass range decoupling after $e^+e^-$  annihilations,
we get
\be
\label{A3}
\hbox{\small$\displaystyle
\frac{\Omega_{dm}\,h^2}{.1}\,\simeq\ \frac{N_{\circ\,dm}\,m_{dm}}{(\rho_c/h^2)\times .1}\, \simeq\,\frac{g}{2}\ \,\frac{m_{dm}}{\hbox{MeV}}\ \
2.03\ 10^5\ (x_F)^{3/2}\ e^{-x_F},
$}
\ee
which determines (e.g. by taking $\,\ln$) $\,x_F$ as a function of $m_{dm}$ (and $g$), e.g. for 1 MeV $\,x_F\simeq 16.4\,$
for $g=2$ (or 17.2\, for $g=4$).

\vspace{2mm}

For heavier particles decoupling after $\,\mu^+\mu^-$ but before $\,e^+e^-$  \,annihilations
there is a further 4/11 reduction factor in the relic density as compared to (\ref{A2}),
so that
\be
\label{A4}
\hbox{\small$\displaystyle
\frac{\Omega_{dm}\,h^2}{.1}\,\simeq\  \frac{g}{2}\ \,\frac{m_{dm}}{\hbox{MeV}}\ \ 7.4\ 10^4\ (x_F)^{3/2}\ \,e^{-x_F} \ .
$}
\ee
This gives approximately, for $\,m_{dm}\,$ = 10 MeV, 100 MeV or 1 GeV,
$\,x_F\,\simeq \,17.8,\ 20.3$ or $22.8$, respectively, for a complex spin-0 or a Majorana particle ($g=2$). For a Dirac particle with $g=4$, these values are increased by \,$\simeq$ .8,
to about \,18.6, 21.1 or 23.6, \,respectively.

\vspace{2mm}

These estimates for $\,x_F$, although na\"{\i}ve, are sufficient
for a first estimate of the required annihilation cross sections at freeze out.
Since we now demand a fixed $\,\Omega_{dm}\,h^2\simeq.1\,$ for any given $\,m_{dm}\,$
(rather than estimating an unknown $\,\Omega_{dm}\,h^2\,$ as a function of $\,m_{dm}\,$
and $\,<\sigma_{ann}\,v_{rel}>$), \,the corresponding $\,x_F\,$ as evaluated above
is directly fixed by  $\,m_{dm}\,$ (and $g$) through (\ref{A3}) or (\ref{A4}),
\,without any direct reference here to $\,<\!\sigma_{ann}v_{rel}\!>$\,
(itself a function of $\,m_{dm}$ \,and $\,x_F$).

\end{document}